\documentclass[12pt,oneside]{article}
\usepackage{amsfonts,amssymb,graphicx}
\def\be{\begin{equation}}
\def\ee{\end{equation}}
\def\bea{\begin{eqnarray}}
\def\eea{\end{eqnarray}}
\def\<{\langle}
\def\>{\rangle}
\def\~{\tilde}
\def\s{\sigma}

\def\a{\alpha}
\def\b{\beta}

\newcommand{\R}{\Bbb R}

\newcommand{\N}{\Bbb N}

\newtheorem{remark}{Remark}

\newtheorem{theorem}{THEOREM}
\newtheorem{definition}{Definition}
\newtheorem{corollary}{Corollary}
\newtheorem{lemma}{Lemma}
\newenvironment{proof}{Proof:}{\hfill$\square$\vskip.5cm}
\begin{document}

\begin{center}{\sc\Large Thermodynamic Limit for \\Mean-Field Spin
Models}
\end{center}

\begin{center}{ A. Bianchi, P. Contucci,
C. Giardin\`a}\\
{\small Dipartimento di Matematica} \\
    {\small Universit\`a di Bologna,
    40127 Bologna, Italy}\\
{\em \{bianchi,contucci,giardina\}@dm.unibo.it}
\end{center}
\vspace{1truecm}
\begin{abstract}\noindent
If the Boltzmann-Gibbs state $\omega_N$ of a mean-field $N$-particle system with Hamiltonian $H_N$ 
verifies the condition
$$
\omega_N(H_N) \; \ge \; \omega_N(H_{N_1}+H_{N_2})
$$
for every decomposition $N_1+N_2=N$, then its free energy density increases
with $N$.
We prove such a condition for a wide class of spin models which includes the Curie-Weiss
model, its p-spin generalizations (for both even and odd p), its random field version and also the finite pattern
Hopfield model. For all these cases the existence of the thermodynamic limit by subadditivity and boundedness
follows.
\end{abstract}
\newpage
\section{Introduction}
The rigorous theory of the thermodynamic limit which already in the sixties was a well established 
part of equilibrium statistical mechanics \cite{Ru} received
recently a new impulse thanks to the treatment of the Sherrington-Kirkpatrick \cite{MPV} model of 
the mean
field spin glass done by Guerra and Toninelli \cite{GuTo}. Moreover in a sequel work \cite{Gu} it became 
clear
that a good control of the limit, especially when it is obtained by monotonicity through subadditivity
arguments, may lead to sharp bounds for the model and carries important informations well beyond the
existence of the limit itself. In this paper we build a theory of thermodynamic limit which apply to a family of
cases including both random and non-random mean field models like the Curie Weiss model \cite{Ba}, its p-spin
generalizations, its random field version \cite{MP}, and also the finite pattern Hopfield model \cite{Ho}. 
Due to the explicit size dependence of the local interactions we stress that mean field models do not
fall into the class for which standard techinques \cite{Ru} can be applyed to prove the existence of
the thermodynamic limit. Moreover even in the Curie Weiss model in which the exact solution is
available it is interesting to obtain the existence of the thermodynamic quantities without
exploiting the exact solution (see \cite{EN,CGI}).

With
respect to the theory relative to the random case \cite{CDGG} we use here a different interpolation technique which
works pointwise with respect to the disorder. The novelty of our approach relies on the fact that while in the
previous case the condition for the existence of the limit is given in terms of a suitably deformed quenched measure,
in the class of models we treat here we are able to give a condition with a direct thermodynamic meaning: the
Bolztmann-Gibbs state for a large system provides a good approximation for the subsystems. The fact that our
existence condition is fully independent from the interpolation parameter relies on the convexity of the
interpolating functional, a property still under investigation for  the spin glass models
\cite{CG1,CG2}.

\section{Definitions and Results.}

We consider a system of N sites: $\{1,2,...,N\}$, to each site we associate 
a spin variable $\s_i$ taking values in $\{\pm 1\}$. A spin configuration is 
specified by the sequence 
$\s = \{\s_1,\s_2,\ldots,\s_N\}$ and we denote the sets of all spin 
configurations by $\Sigma_N=\{\pm 1\}^N$. 
We will study models defined by a {\it mean field} Hamiltonian,
i.e. for a given 
bounded function $g: [-1,1] \rightarrow \R $,
\be
H_N (\s)=-Ng(m_N)
\label{def}
\ee
where
\be
m_N(\s) \, = \, \frac{1}{N}\sum_{i=1}^{N}\s_i \; .
\label{emme}
\ee

\begin{definition}
For each $N$ and a given inverse temperature $\b$ we introduce
the partition function
\be
\label{zeta}
Z_{N}\,=\,\sum_{\s\in\Sigma_N }e^{-\b H_{N}(\s)}\, ,
\ee
the free energy density (and the auxiliary function $\a_{N}$) 
\be
\label{alpha}
-\b f_{N}\,=\,\frac{1}{N}\ln Z_{N}\,=\,\a_{N}\, ,
\ee
and, for a generic  observable $O(\s)$, the Boltzmann-Gibbs state
\be
\omega_{N} (O)\,=\,\frac{\sum_{\s\in\Sigma_N}O(\s)e^{-\b H_N (\s)}}
{Z_{N}} .
\ee
\end{definition}
\begin{remark}
Obviously from (\ref{zeta}) and (\ref{alpha}), one has that if $g$ and $g'$ are two 
functions from $[-1,1]$ to $\mathbb{R}$ such that 
\be
\|g-g'\| := \sup\{|g(x) - g'(x)| : -1\leq x\leq 1\}
\ee
is bounded, then one 
has
\be
  |\alpha_N - \alpha_N'| \leq \beta \|g-g'\|
\ee
for every $N$.
\end{remark}

We can now state our main result:
\begin{theorem}
\label{convergenza}
Let $H_N(\s)$ be a mean field Hamiltonian (see eq. (\ref{def}), (\ref{emme})).
If for every 

partition of the set $\{1,2,...,N\}$ into $\{1,2,...,N_1\}$ and
$\{N_1+1,...,N\}$ with $N=N_1+N_2$ and

\be
H_{N_1}=H_{N_1}(\s_1,...,\s_{N_1}) \; ,\qquad
H_{N_2}=H_{N_2}(\s_{N_1+1},...,\s_{N}) \; ,
\label{h1h2}
\ee
the condition
\be
\label{ridotta}
\omega_{N}(H_N) \; \ge \; \omega_{N}(H_{N_1}+H_{N_2}) \; ,
\ee
is verified, then the thermodynamic limit exists in the sense:
\be
\lim_{N\rightarrow\infty}\a_{N} \;=\;\inf_N \a_{N} \;=\;\a.
\ee
\end{theorem}
\section{Proof}

\begin{definition}
Let us define the interpolating Hamiltonian as a function
of the parameter $t\in[0,1]$:
\be
H_{N}(t)\,=\,tH_{N}+(1-t)[H_{N_{1}}+H_{N_{2}}] \; ,
\label{ht}
\ee
and consider its relative partition function $Z_{N}(t)$, free energy density
$f_N (t)$ and Boltzmann state $\omega_{N,\,t}$.
\end{definition}

\noindent
The interpolation method that we are going to use is based on the sign control
for both the first and second derivative of $\a_{N}(t)$. More precisely the following holds:\\
\begin{lemma}
\label{lemma1}
Let $H_N $ be the mean field Hamiltonian and $H_N (t)$ its relative
interpolation.\\
If 
\be
\frac{d}{dt}\a_N (t)\leq 0
\ee
for all $t\in[0,1]$, then
\be\label{sub}
\a_N \leq \frac{N_1}{N}\a_{N_1} \, + \, \frac{N_2}{N}\a_{N_2} \, ,
\ee 
for each decomposition $N=N_1 +N_2$. 
\end{lemma}
\begin{proof}
trivially follows from the fundamental theorem of calculus
and from the observation that definition (\ref{ht}) implies:
\be
Z_{N}(1)\,=\,Z_{N}\; ,
\ee
\be
\a_{N}(1)\,=\,\a_{N} \; ,
\ee
\be
Z_{N}(0) \; = \;  Z_{N_{1}}Z_{N_{2}} \; ,
\ee
and
\be
\a_{N}(0)\,=\,\frac{N_{1}}{N}\a_{N_{1}}+
\frac{N_{2}}{N}\a_{N_{2}}\;.
\ee 
\end{proof}

\begin{lemma}
\label{d1}
Computing the $t$ derivative of $\a_{N}(t)$, we get:
\begin{eqnarray}
\label{dprima}
\a'(t)=\frac{d}{dt}\frac{1}{N}\log Z_{N}(t)& = &
-\frac{\b}{N}
\sum_{\s\in\Sigma_{N}}[H_{N}-H_{N_{1}} -H_{N_{2}}] \,
\frac{e^{-\b H_{N}(t)}}{Z_{N}(t)}\nonumber\\
& = &-\frac{\b}{N} \omega_{N,\,t}[H_{N}-H_{N_{1}}-H_{N_{2}}].
\end{eqnarray}
%Because to the minus (plus) sign of $\a_N '(t)$, results that $\a_N (t)$
%is a decreasing (increasing) function of $t$; this
%implies an inequality fra $\a_N (0)$ and $\a_N (1)$.
%Then, by remark \ref{valoriestremi}, equations (\ref{sub}) 
%and (\ref{super}) hold.
\end{lemma}

\vspace{0.2cm}

\begin{lemma}
\label{d2}
The second derivative of $\a_{N}(t)$ is positive:
\be
\a_{N}''(t)\,=\,\frac{d^2}{dt^2}\a_N (t)\geq 0\,,
\ee 
\end{lemma}
\begin{proof}
\noindent
a direct computation gives
$$
\a_{N}''(t)= \frac{d}{dt}\left (
-\frac{\b}{N} \omega_{N,\,t} [H_{N}
-H_{N_{1}}-H_{N_{2}}] \right )$$
\be=\frac{\b^{2}}{N} \Big ( \omega_{N,\, t}\left[(H_{N}-H_{N_{1}}
-H_{N_{2}})^{2}\right]-
\omega_{N,\, t}^{2}\,[H_{N}-H_{N_{1}}
-H_{N_{2}}] \Big )\, .
\ee
From Jensen's inequality applied to the convex function $x \mapsto x^2$, 
it follows that $\a_{N}''(t)\geq 0$.
\end{proof}
We are now able to prove the statement of Theorem 1.\\
{\bf Proof of THEOREM 1.}
From Lemma (\ref{d1}) we notice that the hypothesis
(\ref{ridotta}) 
$$
\omega_N \left(H_{N}\right)\geq \omega_N \left(H_{N_{1}} +H_{N_{2}}\right )
$$
is equivalent to the condition
$\a_{N}'(1)\leq 0$. On the other hand 
from Lemma (\ref{d2}) it follows that
$\a_{N}'(t)$ is an increasing function of $t$.
This means that the determination of the sign of $\a_{N}'(t)$ 
can be in general established by the evaluation of the sign in 
the extremes of the interval [0,1].
In particular we have:
\be
\a_{N}'(1)\leq 0\;\Longrightarrow\;\a_{N}'(t)\leq 0,\quad\forall t\in[0,1]
\ee
\\
\noindent
Using now Lemma (\ref{lemma1}), the subadditivity property (\ref{sub}) holds 
for $\a_N$ and then, by standard arguments \cite{Ru},
\be
\lim_{N\rightarrow\infty}\a_N{}\,=\,\inf_N \a_N
\label{tlr}
\ee
The existence of thermodynamic limit finally
follows from boundedness of the function $g$ in
Eq. (\ref{def}). Indeed, calling $K$ the maximum
of $g(x)$ on the interval $[-1,1]$, we have
\be
\a_N = \frac{1}{N} \ln \sum_{\sigma\in \Sigma_N}
e^{\b N g(m_N)} \geq 
\frac{1}{N} \ln e^{\b N K} = \b K \; .
\ee

\section{Applications}

In this Section we identify a class of 
mean field models for which the hypotheses 
of our theorem are verified.
Specifically these will be all models 
such that the function $g$ of formula
(\ref{def}) is convex or polynomial.

\noindent
\begin{corollary} 
\label{coro-convex}
Let the Hamiltonian be of the form
\be
H_N (\s)\,=\,-Ng(m_N)
\ee 
with $g: [-1,1] \rightarrow \R$ a bounded {\em convex} function.
Then the thermodynamic limit of the free energy exists.
\end{corollary}

\noindent
\begin{proof}
For a given $\s \in \Sigma_N$ and for
every decomposition $N=N_1+N_2$ we
define the quantities
\be
\label{parametrispezzati}
m_{N_{1}}(\s)\,=\,\frac{1}{N_{1}}\sum_{i=1}^{N_{1}}\s_{i}\qquad
 m_{N_{2}}(\s)\,=\,\frac{1}{N_{2}}\sum_{i=N_{1}+1}^{N}\s_{i}\;,
\ee
so that the total magnetization is a
convex linear combination of the two:
\be
m_{N}\,=\,\frac{N_{1}}{N}m_{N_{1}}+
\frac{N_{2}}{N}m_{N_{2}}\;.
\ee
Using this definitions the hypothesis
(\ref{ridotta}) of Theorem (\ref{convergenza})
is verified:
\be
\omega_N \left(H_N -H_{N_1}-H_{N_2}\right)=-N\omega_N
 \left(g(m_{N})-\frac{N_{1}}{N}g(m_{N_{1}})-\frac{N_{2}}{N}g(m_{N_{2}})
\right)\; \geq 0
\ee
where the last inequality follows from convexity of $g$.
\end{proof}

\begin{remark}
\label{piuvariabili}
The previous Corollary can be obviously generalized 
to the case where the function $g$ is a {\em convex}
bounded function of many variables, each
of them fulfilling the property (\ref{parametrispezzati}).
\end{remark}

\begin{corollary} 
\label{coro-poly}
Let the Hamiltonian be of the form
\be
H_N (\s)\,=\,-Ng(m_N)\,,
\ee 
with $g: [-1,1] \rightarrow \R$ a {\em polynomial} function
of degree $n\in \N$.
Then the thermodynamic limit exists.
\end{corollary}

\noindent
\begin{proof}
First of all we consider the case $g(x)=x^k$ (the generalization 
will follow in a simple way) with
associated Hamiltonian 
\be\label{uno}
H_N (\s)\,=\,-N m_N^k\,=\,-\frac{1}{N^{k-1}}\sum_{i_1,i_2,\ldots,i_k =1}^{N}
\s_{i_1}\s_{i_2} \cdots \s_{i_k}\,.
\ee
By splitting  the summation into two pieces,
the first containing the summation with indexes 
all different among themselves,
the second containing the remaining terms, 
we have
\be
H_N (\s)\,=-\frac{1}{N^{k-1}} \left [
\sum_{i_1\ne \ldots\ne i_k}\s_{i_1} \cdots \s_{i_k}
+  \sum_{i_1,\ldots,i_k}^{\ast}\s_{i_1}\cdots \s_{i_k} \right ]
\ee
where the second summation $\sum^{\ast}$
includes all terms with at least two equal 
indexes. A simple computation shows that
\be
\frac{1}{N^{k-1}} \sum_{i_1,\ldots,i_k}^{\ast}\s_{i_1}\cdots \s_{i_k}\,= \,{\cal O}(1)
\ee
Defining now the model with Hamiltonian
\be\label{due}
\tilde{H}_N (\s)= -\frac{1}{(N-1)(N-2)\cdots (N-k+1)}\sum_{i_1 \ne i_2
 \ne \ldots\ne i_k = 1}^N \s_{i_1} \s_{i_2} \cdots \s_{i_k}\,,
\ee
it follows that
\bea\label{tre}
H_N (\s)& = & \frac{(N-1)(N-2)\cdots (N-k+1)}{N^{k-1}}
                 \tilde{H}_N (\s) + {\cal O}(1) \nonumber\\
        & = & \tilde{H}_N (\s) + {\cal O}(1) 
\eea  
Using Remark 1 one has
that the two models $H_N$ and $\tilde{H}_N$ have the
same thermodynamic limit (if any).
On the other hand for the model $\tilde{H}_N$ 
we have
\bea
\omega_N (\tilde{H}_N )\, & = & \,-\frac{1}{(N-1)(N-2)\cdots (N-k+1)}
 \sum_{i_1 \ne i_2 \ne \ldots\ne i_k}\omega_N(\s_{i_1} \s_{i_2} \cdots \s_{i_k})
\nonumber \\
& = & \,-N\omega_N (\s_1 \s_2 \cdots \s_k)\,
\eea 
where the last equality follows from permutation invariance
( $\omega_N (\s_{i_1} \s_{i_2} \cdots \s_{i_k})$
does not depend on the choice of the indexes).
Analogously one can repeat the same computation for $\tilde{H}_{N_1}$
and  $\tilde{H}_{N_2}$ in the state $\omega_N$. By permutation symmetry, this yields that 
Hypothesis (\ref{ridotta}) of Theorem (\ref{convergenza})
is verified as an equality
\be
\omega_N \left(\tilde{H}_N -\tilde{H}_{N_1}-\tilde{H}_{N_2}\right)= 0\,,
\ee
implying the existence of thermodynamic limit for the model $\tilde{H}_N$,
and so for the model $H_N$.

\noindent
Since we have proved the Corollary for $g(x) = x^k$,
the case of a generic polynomial function of degree $n$ 
\be
\label{polinomio}
g(x)=\sum_{k=0}^{n}a_k x^k
\ee
is treated by the same argument,
applied to each monomial of the sum.
\end{proof}

\begin{corollary}
Using Remark 1 and the 
Stone-Weierstrass theorem, the thermodynamic limit of $\alpha_N$ exists 
if, instead of $g$ being polynomial, $g$ is merely continuous up to the 
boundary of $[-1,1]$.
\end{corollary}

\subsection{Examples}
\begin{enumerate}
\item {\bf The Curie-Weiss models.}\\
For every  integer $p$, with $p < N$,
consider the model defined by
\be 
\label{HCW}
 H_{N}(\s)\,
:=\,-\frac{1}{N^{p-1}}\sum_{i_1,i_2,\ldots,i_{p}=1}^N
\s_{i_1} \s_{i_2} \cdots \s_{i_{p}}
\ee 
which represents the generalized
Curie Weiss model with $p$-spin interaction.
The standard Curie-Weiss model corresponds
to the case $p=2$.
From eq. (\ref{emme}) the previous Hamiltonian can be written as
\be
\label{HCCW}
H_{N}\,=\,-N m_N^{p}
\ee

and the existence of the thermodynamic limit is
then implied by Corollary (\ref{coro-poly}).
Moreover, the same result holds for any linear 
combination (see Eq.(\ref{polinomio})) of generalized Curie Weiss models 
with $p$-spin interaction, both ferromagnetic 
and antiferromagnetic.

\item
{\bf The random field Curie Weiss model.}

Here we consider the model defined by (see \cite{MP} for a review)
\be
\label{rfcw}
H_{N}(h,\s):=-\frac{1}{N}
\sum_{i,j=1}^{N}\s_{i}\s_{j} + \sum_{i=1}^{N} h_i \s_i
\ee
where $\{h_i\}_{i=1,\ldots,N}$ is a family of i.i.d.
Bernoulli random variables, with probability distribution
\be
p(h_i) = \left\{ \begin{array}{ll}
1/2, & \mbox{if $h_i = 1$}, \\
1/2, & \mbox{if $h_i = -1$}.       
\end{array}
\right.
\ee  
For a given realization of the random field $h$,
we define the quantities
\be
m_N^{+}(\s,h) = \frac{1}{N}\sum_{i=1}^{N} 
\frac{1+h_i}{2}\,\s_i
\ee
\be
m_N^{-}(\s,h) = \frac{1}{N}\sum_{i=1}^{N} 
\frac{1-h_i}{2}\,\s_i
\ee
The Hamiltonian can be written in terms of
these variables as
\be
H_N = - N g(m_N^{+} , m_N^{-})
\ee
where 
\be
g(m_N^{+} , m_N^{-}) = (m_N^{+} + m_N^{-})^2 - (m_N^{+} - m_N^{-})
\ee
Since this function is obviously convex with respect to
both $m_N^{+}$ and $m_N^{-}$, bounded by $2$, and 
\be
m_{N}^{\pm}(\s,h)\,=\,\frac{N_{1}}{N}m_{N_{1}}^{\pm}(\s,h)+
\frac{N_{2}}{N}m_{N_{2}}^{\pm}(\s,h)\;
\ee 
the Corollary (\ref{coro-convex})
can be applied and we find
(pointwise in the $h$'s)
\be
\a_N(h) \, \le \, \frac{N_1}{N}\a_{N_1}(h) +  \frac{N_2}{N}\a_{N_2}(h) \; . 
\ee
Averaging now over the $h$'s, the subadditivity
property for quenched $\a_N$ is proved and this
yields (\ref{tlr}).

\item 
{ \bf The Hopfield model.}

\noindent
The Hamiltonian of the Hopfield model (see \cite{Bo} for a review)
is given by :
\be 
\label{HopH}
H_{N}(\xi,\s):=-\sum_{\mu=1}^{M}\frac{1}{N}
\sum_{i,j=1}^{N}\xi_{i}^{\mu}\xi_{j}^{\mu}\s_{i}\s_{j}
\ee
where $M$ is the (fixed) number of pattern and the 
$\{\xi_i^{\mu}\}_{i=1,\ldots,N}^{\mu = 1,\ldots,M}$
is a family of i.i.d. Bernoulli variables
with probability distribution
\be
p(\xi_i^{\mu}) = \left\{ \begin{array}{ll}
1/2, & \mbox{if $\xi = 1$}, \\
1/2, & \mbox{if $\xi = -1$}.       
\end{array}
\right.
\ee
Defining the quantities 
\be
\label{parametrihop}
m_{N}^{\mu}(\s,\xi)=\frac{1}{N}\sum_{i=1}^{N}\xi_{i}^{\mu}\s_{i}\quad,
\forall\mu=1,\ldots,M\;,
\ee
the Hamiltonian (\ref{HopH}) can be written as
\be \label{HopH2}
H_{N}(\s,\xi)=-N\sum_{\mu=1}^{M}(m_{N}^{\mu}(\s,\xi))^{2}.
\ee

The model can be included in the general
treatment of the previous section by
considering a function $g$ of $M$ variables,
%all of which are additive in system size
\be
g(m_N^1(\s,\xi),\ldots,m_N^M(\s,\xi)) = \sum_{\mu=1}^M (m_N^{\mu}(\s,\xi))^2
\ee
such that
\be
H_{N} \, = \, -N  g(m_N^1,\ldots,m_N^M)  .
\ee
Since
\be
m_{N}^{\mu}(\s,\xi)\,=\,\frac{N_{1}}{N}m_{N_{1}}^{\mu}(\s,\xi)+
\frac{N_{2}}{N}m_{N_{2}}^{\mu}(\s,\xi)\;\quad\quad \forall \mu=1,\ldots,M
\ee
and the function $g$ is convex with respect to
every $m_N^{\mu}$ and bounded by $M$,
using Corollary (\ref{coro-convex})
we have (pointwise in the $\xi$'s)
\be
\a_N(\xi) \, \le \, \frac{N_1}{N}\a_{N_1}(\xi) +  \frac{N_2}{N}\a_{N_2}(\xi) \; . 
\ee
Averaging over the $\xi$'s yields the (\ref{tlr}).

\begin{remark}
\label{nongen}
We want to notice that the method shown doesn't apply to
the Hopfield model with a thermodynamically growing number of 
patterns defined for every positive constant $\gamma$ by
\be 
\label{Hopinf}
H_{N}(\xi,\s):=-\sum_{\mu=1}^{\gamma N}\frac{1}{N}
\sum_{i,j=1}^{N}\xi_{i}^{\mu}\xi_{j}^{\mu}\s_{i}\s_{j} \; ,
\ee
because this Hamiltonian is not of the form (\ref{def}).
\end{remark}

\end{enumerate}

\vskip .6 cm
\noindent
\noindent {\large \bf Acknowledgments.}
C.G. thanks A. Bovier for helpful correspondence.
P.C. thanks A. Bovier, F.L.Toninelli and F. Guerra for many 
enlightening discussions. We thank S. Graffi for his
continuous encouragement and advice. We are grateful to the referee
for several helpful comments in particular for the suggestion of Corollary 3.

\end{document}